# Blind Deconvolution of Ultrasonic Signals Using High-Order Spectral Analysis and Wavelets

Roberto H. Herrera[1], Eduardo Moreno[2], Héctor Calas[2], and Rubén Orozco[3]

[1] University of Cienfuegos, Cuatro Caminos, Cienfuegos, Cuba
henry@finf.ucf.edu.cu
[2] Institute of Cybernetics, Mathematics and Physics (ICIMAF), Havana, Cuba
{moreno, hcalas}@icimaf.inf.cu
[3] Central University of Las Villas, Santa Clara, Cuba
rorozco@fie.uclv.edu.cu

**Abstract.** Defect detection by ultrasonic method is limited by the pulse width. Resolution can be improved through a deconvolution process with a priori information of the pulse or by its estimation. In this paper a regularization of the Wiener filter using wavelet shrinkage is presented for the estimation of the reflectivity function. The final result shows an improved signal to noise ratio with better axial resolution.

## 1 Introduction

Deconvolution of ultrasonic signals is defined as the solution of the inverse problem of convolving an input signal, known as the transducer impulse response $h(n)$ and medium reflectivity function $x(n)$ and can be represented by [1]:

$$y(n) = h(n) * x(n) + \eta(n) \; . \tag{1}$$

where $y(n)$ is the measured signal, the operator * denotes the convolution operation and $\eta(n)$ is the additive noise. To recover $x(n)$ from the observation $y(n)$ drives to improve the appearance and the axial resolution of the images through the elimination of the dependent effects of the measuring system [1]. The signal $y(n)$ corresponds to A-scan lines of 2-D acoustic image or 1-D signal, where the problem settles down by taking the desired signal $x(n)$ as the input of a linear time invariant system (LTI) with impulse response $h(n)$ [2]. The output of the LTI system is blurred by white Gaussian noise $\eta(n)$ of variance $\sigma^2$. In frequency domain from (1) we get:

$$Y(f) = H(f)X(f) + N(f) \; . \tag{2}$$

Where: $Y(f)$, $H(f)$ and $N(f)$ are the Fourier Transform of $y(n)$, $h(n)$ y $\eta(n)$ respectively. If the system frequency response $H(f)$ does not contain zeros an estimation of $x(n)$ can be obtained from:

$$X_1(f) = H^{-1}(f)Y(f) = X(f) + H^{-1}(f)N(f) \; . \tag{3}$$

However where $H(f)$ takes near to zero values, the noise is highly amplified with variance spreading to infinite which leads to incorrect estimates. In this case it is necessary to include in the inverse filter some regularization parameter which reduces the variance of the estimated signal. The most known case of regularized filter for stationary signals is the Wiener filter [3].

When the signals under analysis shows non stationary properties, as abrupt changes, the Wiener filter based on the Fourier Transform does not give satisfactory results in the estimation, conditioned by the characteristics of Fourier basis ($e^{jw}$) [1]. A projection into a base that can characterize these non stationary signals and at the same time achieves a better matching with the transmitted pulse, as wavelets, drives to a better localization in time and frequency [3]. Another of the advantages of wavelets is that the signals can be represented with some few coefficients different from zero, what corresponds with the ultrasonic signals, where the trace is only composed by values different from zero in cases of abrupt changes of acoustic impedance, this leads to an efficient methods of compression and noise filtering. R. Neelamani, H. Choi & R. Baranikuk, recently proposed a regularized deconvolution technique based on Wavelet (ForWaRD) [4] which will be used in this paper for the deconvolution of ultrasonic signals as a first step to the conformation of acoustic images by means of Synthetic Aperture Focusing Testing (SAFT).

The initial problem in deconvolution, is the a priori knowledge or not of the system impulse response *h(n)*. Oppenhiem & Shafer have defined the case of estimating *x(n)* from *h(n)* as the well-known homomorphic deconvolution [5], using the real cepstrum for minimum phase signals or the complex cepstrum for the most general case. Another author, Torfinn Taxt in [6], compares seven methods based on the cepstrum for blind deconvolution (without knowing *h(n)*), in the estimation of the reflectivity function in biological media. We select the method of High Order Spectral Analysis (HOSA) because of its immunity to the noise and the not initial conditionality that the transducer's electromechanical impulse response is of minimum phase, something that depends on the construction of the housing of the pieso-electric and of the impedance matching between the transmitter and the ceramic [7].

The paper is structured as follows. Section 2.1 deals with the process of estimating the system function using HOSA. Section 2.2 summarizes the procedure for a first estimate using the Wiener filter. Section 2.3 focuses on the wavelet-based regularized deconvolution. Section 2.4 describes the measurement system and the signals to be processed. Section 3 presents the results with a comparative analysis. Finally Section 4 gives the conclusions of the paper.

## 2   Materials and Methods

This Section firstly describes the method used for estimating the system function and continues with the wavelet-based deconvolution.

### 2.1   Estimation of System Function Using HOSA

The system function described in (1) as the transducer's impulse response *h(n)* is a deterministic and causal FIR filter, *x(n)* represents the medium response function that we assume initially, without loss of generality, stationary, zero mean and non Gaus-sian distribution, this last property guarantees that its third-order cumulant exists, like we will explain later on, on the other hand $\eta(n)$ represents the zero mean Gaussian noise that is uncorrelated with $x(n)$. The third-order cumulant of the zero mean signal $y(n)$ is represented by [1], [8]:

$$c_y(m_1, m_2) = \gamma_x \frac{1}{M} \sum_{k=0}^{M-1} h(k)h(k+m_1)h(k+m_2) \ . \tag{4}$$

where $\gamma_x = E[x^3(n)]$, is a constant equal to the third cumulant of the signal $x(n)$, and E is the operator of statistical average.

By applying the 2-D Z-Transform ($Z_{2D}$) to (3) we get the bispectrum:

$$C_y(z_1, z_2) = \gamma_x H(z_1) H(z_2) H(z_1^{-1} z_2^{-1}) \ . \tag{5}$$

The bicepstrum $by(m1, m2)$, is obtained as was described in [5], logarithm of the bispectrum and inverse transformation to arrive into the 2-D quefrency domain:

$$b_y(m_1, m_2) = Z_{2D}^{-1} \left[ \log(C_y(z_1, z_2)) \right] \ . \tag{6}$$

As follows in [1], the cepstrum $\hat{h}(n)$ of $h(n)$ is obtained by evaluating the bicepstrum along the diagonal $m1 = m2$ for all $n \neq 0$:

$$\hat{h}(n) = b_y(-n, n) \quad \forall n \neq 0 \ . \tag{7}$$

Then from (6) we can estimate $h(n)$ as:

$$h(n) = Z^{-1} \left\{ \exp\left[ Z(\hat{h}(n)) \right] \right\} \ . \tag{8}$$

where $Z$ and $Z^{-1}$ are 1-D the direct and inverse Z-Transform respectively.

The bicepstrum is derived from the bispectrum in the same way that the cepstrum is obtained from the spectrum. The main advantage of this estimation method is that the bispectrum of the white Gaussian noise is zero [7], which allows us to estimate $h(n)$ without taking into account the contribution of $\eta(n)$ in (1).

## 2.2 The Wiener Filter

Having $h(n)$ we can estimate $X_1(f)$ using the Wiener filter:

$$X_1(f) = Y(f) \left[ \frac{H^*(f)}{|H(f)|^2 + q} \right] \ . \tag{9}$$

Where $q$ is a term that includes the regularization parameter and the noise contribution, $H(f)$ is the 1-D Fourier Transform of $h(n)$ and $H^*(f)$ its complex conjugated, the term inside the brackets is the inverse Wiener filter in generic form it is represented by [1]:

$$G(f) = \frac{H^*(f)P_{x_1}(f)}{|H(f)|^2 P_{x_1}(f) + \alpha\sigma^2} \ . \tag{10}$$

where $Px_1(f)$, is the power spectral density of $x_1(n)$, $\alpha$ is the regularization parameter and $\sigma^2$ represents the noise variance. As $Px_1(f)$ is unknown it is necessary to use the iterative Wiener method, in this study we took $\alpha$=0.01 initially, giving good results in the estimate and $\sigma^2$ was calculated as the median of the finest scale wavelets coefficients of $y(n)$ [8], $x_1(n)$ is obtained from $X_1(f)$ by inverse Fourier transformation.

### 2.3 Wavelet-Based Wiener Filter

The discrete wavelet transform (DWT) represents a 1-D continuous-time signal $x(t)$, in terms of shifted versions of a lowpass scaling function $\phi$ and shifted and dilated versions of a prototype band-pass wavelet function $\psi$ [4]. As it was demonstrated by I. Daubechies [9], special cases of these functions $\psi_{j,k}(t) = 2^{j/2}\psi(2^j t - k)$ and $\psi_{j,k}(t) = 2^{j/2}\psi(2^j t - k)$ form an orthonormal basis in the $L^2(\Re)$ space, with $j,k \in \mathbb{Z}$. The parameter $j$ is associated with the scale of the analysis and $k$ with the localization or displacement. Signal decomposition at a level $J$, would be given by [1]:

$$x^J(t) = \sum_{k=1}^{2^{(N-J)}} c(k)\phi_k(t) + \sum_{j=1}^{J}\sum_{k=1}^{2^{(N-J)}} d(j,k)\psi_{j,k}(t) \ . \tag{11}$$

where $c(k)$ is the inner product $c(k) = \langle x(t), \phi_{j,k}(t) \rangle$ and $d_{j,k} = \langle x(t), \psi_{j,k}(t) \rangle$.

The estimated signal from the Wiener filter is projected into this base, and at each decomposition level the variance $\sigma_j^2$ is obtained for noise reduction. The following step is to use the Wiener filter in the wavelet domain where the filtering process is done for the wavelet coefficients. From (10) we have [4]:

$$\lambda_{j,k}^d = \frac{|d_{j,k}|^2}{|d_{j,k}|^2 + \sigma_j^2} \text{ and } \lambda_{j,k}^c = \frac{|c_k|^2}{|c_k|^2 + \sigma_j^2} \ . \tag{12}$$

By substituting (11) in (10) we obtain the expression of the estimated reflectivity function $\tilde{x}(n)$:

$$\tilde{x}(n) = \sum_{k=1}^{2^{(N-J)}} \lambda_{j,k}^c c(k)\phi_k(n) + \sum_{j=1}^{J}\sum_{k=1}^{2^{(N-J)}} \lambda_{j,k}^d d_{j,k}\psi_{j,k}(n) \ . \tag{13}$$

### 2.4 Experimental Setup

The experimental system consisted on the obtaining an acoustic image of 10 bars of acrylic of diameter 5 mm, submerged in water. A data set of 400 RF-sequences has been generated, each RF-sequence containing 9995 sampling points. The RF-lines were sampled at a rate of 50 MHz. An unfocused 3.5 MHz transducer was used in both emission and reception operating in pulse-echo mounted in a scanner controlled by stepping motor with 0.25 mm between A-scan lines.

## 3 Results and Discussion

The deconvolution process steps, as has been described previously include:

1. Estimate the impulse response from the bicepstrum.
2. Obtain a first estimate of the reflectivity function using the regularized Wiener filter in the domain of the frequency.
3. Apply a noise filtering over the wavelets coefficients.
4. Estimate the reflectivity function with the Wiener filter in the wavelet domain.

### 3.1 Estimation of the Ultrasound Pulse

The pulse estimation was carried out on a set of 16 zero mean signals. Fig. 1 shows the obtained pulse, using the MatLab® function bicepsf.m of the HOSA Toolbox.

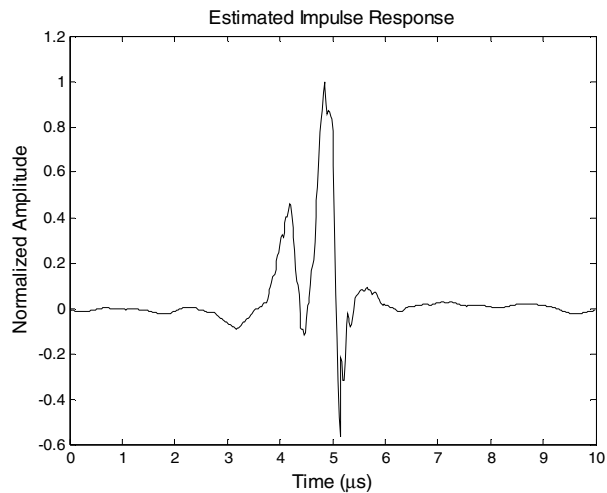

**Fig. 1.** Estimated impulse response. The normalized pulse width *vs* time in μs.

The spectral content of the obtained pulse includes the same band of the original signal.

### 3.2 Estimation of the Reflectivity Function

We used an iterative Wiener filter to estimate the power spectral density $Px_1(f)$, as was explained in the section 2.2. After ten iterations the signal $x_1(n)$ was obtained. Fig. 2 shows a segment of the original signal and the estimated one.

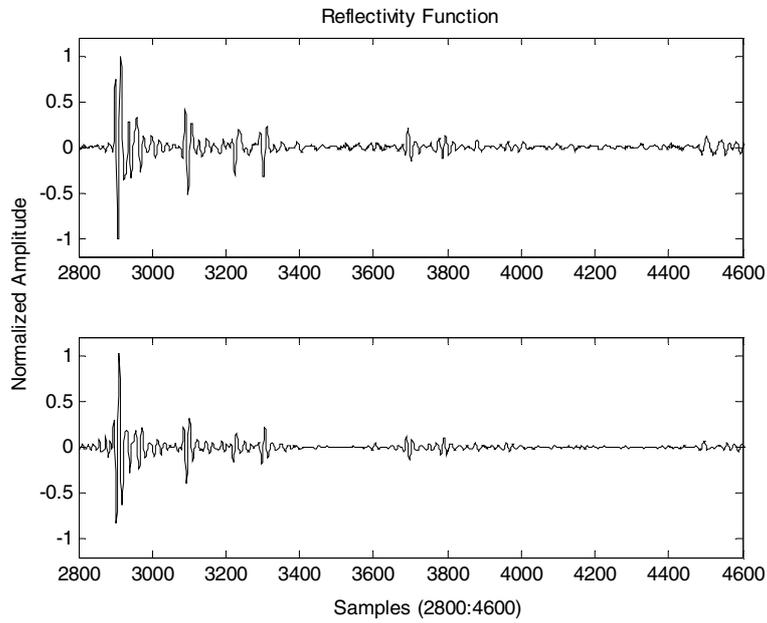

**Fig. 2.** Estimated reflectivity function obtained by iterative Wiener filter. (upper plot) The original signal *y(n)*; (lower plot) the estimated signal $x_1(n)$.

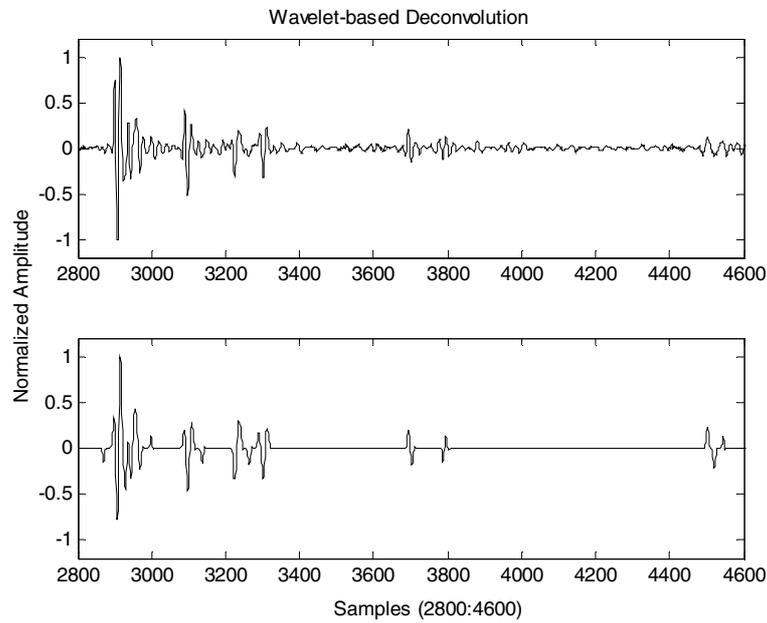

**Fig. 3.** The Wiener filter applied to the wavelets coefficients. (upper plot) The original signal *y(n)*; (lower plot) the deconvolved signal $\tilde{x}(n)$.

### 3.3 Noise Filtering

We used a soft threshold over the wavelets coefficients after a decomposition using DB16 and DB10 in the algorithm proposed in [4]. Fig. 3 shows the result of the deconvolution in the wavelet domain.

The estimated signal shows a better spatial localization, which improves the axial resolution.

In accordance with Fig. 4, the deconvolution of the RF signal improves the resolution, quantified as the decrease of the main lobe width of the autocovariance function [7]. The lobe width at half amplitude (-6dB drop) given in samples is 9 samples for the original signal and 4 for estimated one.

We obtained an increment of the axial resolution in a factor of 2.25. The same procedure was applied to the set of 30 signals of a total of 400 to characterize the standard deviation of the values, obtaining a factor of 2.25± 0.36.

This increment of the axial resolution depends of the transducer's spectral properties; consequently it is suggested to prove the method with different frequency bandwidth ratio.

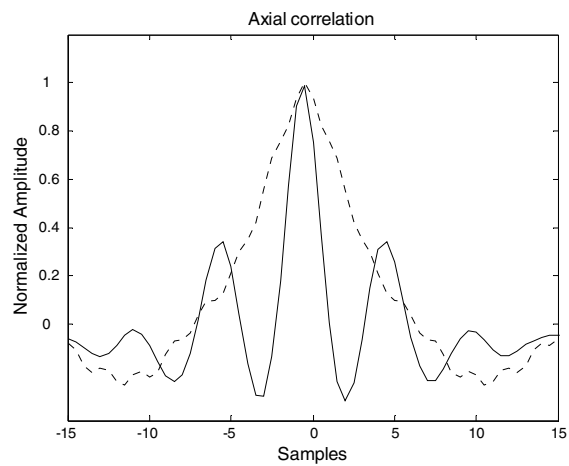

**Fig. 4.** Autocovariance function of the original signal (*dotted line*) and of the estimated signal (*continuous line*). The amplitude was normalized in both functions and centered in their maximum.

## 4   Conclusions

This paper establishes a cepstrum-based method using high-order statistics as the first step for the blind deconvolution kernel estimation which is used in the inverse filter design in both Fourier and wavelet domain for the reconstruction of the medium reflectivity function. This procedure results in a significant reduction of the time spatial support, suggesting a significant gain in the axial resolution.

This property is particularly useful in the case of acoustic image generation, where we will apply these results.